\documentclass[prb,twocolumn]{revtex4}
\usepackage{amsmath}
\usepackage{graphicx}
\usepackage{bbold}

\newcommand{\spinor}[1]{#1}

\newcommand{\ddd}{\mathrm{d}}

\newcommand{\ddif}[2]{\frac{\ddd #1}{\ddd #2}}

\newcommand{\ccr}[1]{c^{\dagger}_{#1}}

\newcommand{\comm}[1]{\left[#1\right]}

\newcommand{\ket}[1]{\left| {#1}\right\rangle}

\newcommand{\spup}{\uparrow}
\newcommand{\spdown}{\downarrow}

\begin{document}

\title{Rate equations for Coulomb blockade with ferromagnetic leads}
\author{Stephan Braig and Piet W. Brouwer}
\affiliation{Laboratory of Atomic and Solid State Physics,
Cornell University, Ithaca, NY 14853
}
\date{\today}

\begin{abstract}
We present a density-matrix rate-equation approach to sequential
tunneling through a
metal particle weakly coupled to ferromagnetic leads. The
density-matrix description is able to deal with correlations
between degenerate many-electron states that the standard rate
equation formalism in terms of occupation probabilities cannot
describe. Our formalism is
valid for an arbitrary number of electrons on the dot, for
an arbitrary angle between the polarization directions of
the leads, and with or without spin-orbit scattering on the metal
particle. Interestingly, we find that the density-matrix description
may be necessary even for metal particles with unpolarized leads if
three or more single-electron levels contribute to the transport
current and electron-electron interactions in the metal particle are
described by the `universal interaction Hamiltonian'.
\end{abstract}
\maketitle

\section{Introduction}

Spin-polarized electron tunneling is essential to spin-based
electronics~\cite{kn:zutic2004} and nanoscale magnetics based on the
spin-transfer effect.~\cite{kn:berger1996,kn:slonczewski1996} Whereas
tunneling through a single tunnel barrier, either between two
ferromagnets or between a ferromagnet and a normal metal, has been
studied since the mid 1970s,~\cite{kn:julliere1975} the study of
spin-polarized transport through mesoscopic double tunneling junctions
is more recent. Double mesoscopic junctions are of interest because of
the small capacitance of the central region in between the tunneling
junctions, which allows electrons to be transported one by
one.~\cite{kn:grabert1992} Experiments have been reported both for
normal metal leads with a ferromagnetic island between the
junctions~\cite{kn:gueron1999,kn:deshmukh2001} and for a normal island
with spin-polarized
junctions.~\cite{kn:deshmukh2002,kn:potok2003,kn:pasupathy2004} A large number of theoretical works has dealt with these cases.~\cite{kn:barnas1998,kn:takahashi1998,kn:brataas1999,kn:barnas2000,kn:rudzinski2001,kn:usaj2001,kn:martinek2002,kn:sergueev2002,kn:fransson2002,kn:cottet2004,kn:cottet2004b,kn:ma2003,kn:martinek2003,kn:martinek2003a,kn:choi2004,kn:martinek2004,kn:pedersen2004,kn:waintal2003,kn:koenig2003,kn:braun2004}

In
this work, we consider the case of a normal metal island with ferromagnetic leads.
If the temperature is much larger than the tunneling rates onto or off
the island, electron tunneling is sequential. In that case, quantum
mechanical correlations between electrons in different states are lost
because of thermal smearing, and a simple description in terms of rate
equations applies. Depending on whether the temperature is small
or large compared to the level spacing in the island, these rate
equations describe the probability to find a certain 
number of electrons on the
island~\cite{kn:shekhter1972,kn:kulik1975,kn:averin1986} or the
occupation of the electronic states in the
island.~\cite{kn:beenakker1991,kn:averin1991} 

Whereas the
rate-equation approach 
was applied straightforwardly to 
ferromagnetic leads with collinear 
polarizations,~\cite{kn:waintal2003,kn:cottet2004} application
to leads with non-collinear polarizations requires a formulation
in terms of the density matrices of degenerate levels, not the 
occupations of states.~\cite{kn:koenig2003,kn:braun2004} (If spin degeneracy on the normal metal island is lifted, e.g., by a magnetic field, scalar rate equations remain valid despite polarized leads.)
There are two reasons for this additional
complication: First, with non-collinear polarizations,
no common quantization direction exists, and one cannot avoid
a formulation of the problem in which electrons tunnel into 
superpositions of states with different spin
projections.~\cite{kn:usaj2001} Second,
coupling to the ferromagnetic leads slightly lifts the spin
degeneracy and leads to a slow precession of the spin on the
dot.~\cite{kn:koenig2003} The use 
of density matrices instead of
occupation probabilities in the rate equation formalism
allows for the inclusion of correlations between different
quantum states.~\cite{kn:nazarov1993,kn:gurvitz1998} Since the temperature is much larger than the escape 
rate to the leads, only correlations between states with the
same energy need to be 
taken into account.

Density-matrix rate equations were first used to describe
transport through a 
metal particle (or a quantum dot or a single molecule) with
spin-polarized tunnel contacts in a recent paper by K\"onig 
and Martinek~\cite{kn:koenig2003} (see also Ref.~\onlinecite{kn:braun2004}). These authors used the Keldysh
formalism to derive the density-matrix rate equations for a dot 
in which only one level contributes to transport. The purpose 
of the present work is to formulate
a density-matrix rate equation for quantum dots in which many
electronic levels contribute to transport and to 
simplify the derivation
of Ref.~\onlinecite{kn:braun2004}. The extension to many
levels is relevant
for the analysis of 
 experimental data, since the majority of
 experiments feature high bias voltages at which more than
one electronic level contributes to the current.~\cite{kn:vondelft2001} 

A remarkable result of our study is that a formulation in terms of
density-matrix rate equations is not only needed for
spin-polarized leads with non-collinear polarization directions, 
but that it may also be necessary for unpolarized leads or for
spin-polarized leads with collinear polarization directions if the
metal island has a large dimensionless
conductance $g$. These relatively large metal particles or quantum dots
have degenerate or almost-degenerate many-electron levels. 
Correlations between the degenerate states 
persist during the time an electron occupies the quantum dot
and need to be accounted for using a description in terms of a
density matrix. The origin of the degeneracy is that in large-$g$
metal grains or quantum dots electronic interactions are described
by the `universal interaction Hamiltonian'.\cite{kn:aleiner2002}
With this interaction, many-electron levels with three
or more singly occupied single-electron levels are degenerate if 
their spin is not maximal. For example, there are four degenerate 
states with three singly occupied levels and total spin $S=1/2$. 
A rate equation in terms 
of scalar occupation probabilities only\cite{kn:alhassid2002} 
will fail to describe correlations between these degenerate states. 
A detailed description of this case will be given in Sec.~\ref{sec:unpolarized} below.

\section{Matrix Rate-Equation Formalism}
\label{sec:2}

\begin{figure}[t]
\setlength{\unitlength}{1cm}
\begin{picture}
(5,3.5)(0,0) \put(0,0){\includegraphics[width=5cm]{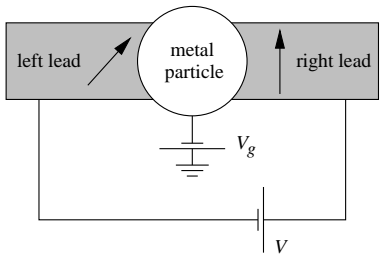}}
\end{picture}
\caption{Metal particle attached to two ferromagnetic leads. 
\label{fig:doublelead}}
\end{figure}

We consider a metal particle --- or a quantum dot or a single
molecule --- that is attached to a number of 
ferromagnetic leads via tunneling contacts with a conductance
much smaller than the conductance quantum $e^2/h$. 
A schematic drawing of a metal particle with two leads is shown
in Fig.~\ref{fig:doublelead}.
In our formulation of the problem, we assume that all leads 
are fully polarized; a partially polarized lead is simply
represented by two fully polarized leads with different densities of
states and different tunneling rates. 
We assume that the temperature $T$ is much larger than the 
tunneling rates to and from the leads. 
This is the regime 
for which rate equations have 
been shown to be a valid description of metal particles without
spin-polarized leads. 

\subsection{Single doubly degenerate level}

In order to make the connection to previous works,\cite{kn:usaj2001,kn:koenig2003,kn:braun2004} we first develop the formalism for the case of nonlinear 
transport through a single level. Our approach is closely connected to the works by Nazarov~\cite{kn:nazarov1993} and Gurvitz,~\cite{kn:gurvitz1998} who used rate equations to describe high-bias transport through a sequence of tunnel barriers.
For a metal particle in which only one level is relevant for transport,
we need to consider occupation of the level
by zero, one, or two electrons, at energies $\varepsilon_0$, 
$\varepsilon_1$, and $\varepsilon_2$, respectively. The precise
value of these energies depends on the charging energy and
exchange interaction of the metal particle, the voltages on 
nearby gates, etc. For occupation by zero or two
electrons, the many-electron level is non-degenerate, and we can
use scalars $p_0$ and $p_2$ to describe the probability
to find the particle in a state with zero or two electrons, respectively. If the
level is occupied by one electron only, one needs to use a $2 \times
2$ density matrix $\rho_1$ to fully describe the state of the particle.
Conservation of probability implies 
\begin{equation}
  p_0 + \mbox{tr}\, \rho_1 + p_2 = 1.
  \label{eq:prob}
\end{equation}

Without tunneling to and from the reservoirs, $p_0$, $\rho_1$, 
and $p_2$ are time independent. The time dependence of 
$p_0$, $\rho_1$, and $p_2$ then arises from tunneling of 
electrons onto or off the metal particle: real
 tunneling processes
shift the number of electrons on the metal particle, whereas
virtual tunneling processes cause a precession of the spin if
the level is singly occupied. The net tunneling rate to and from
lead $\alpha$ depends on the direction of the polarization in
that lead, which we describe by means of the spinors 
$\spinor{m}_{\alpha}$ ($\bar{\spinor{m}}_{\alpha}$) pointing
parallel (anti-parallel) to the polarization direction of 
lead $\alpha$, the tunneling rate $\Gamma_{\alpha}$ for electrons 
with spin $\spinor{m}_{\alpha}$, and the distribution function 
$f_{\alpha}$ in lead $\alpha$. In order to describe both the
virtual and real tunneling processes, we combine the rate 
$\Gamma_{\alpha}$ and the spinors $\spinor{m}_{\alpha}$ and 
$\bar{\spinor{m}}_{\alpha}$ into the spinor tunneling amplitudes 
\begin{equation}
  \gamma_{\alpha} = \Gamma_{\alpha}^{1/2} \spinor{m}_{\alpha},
   \ \
  \bar{\gamma}_{\alpha} = \Gamma_{\alpha}^{1/2} 
  \bar{\spinor{m}}_{\alpha},
\end{equation}
with $\gamma_\alpha^\dagger\gamma_\alpha^{\vphantom{\dagger}}=\bar{\gamma}_\alpha^\dagger\bar{\gamma}_\alpha^{\vphantom{\dagger}}=\Gamma_\alpha$ and $\gamma_\alpha^\dagger\bar{\gamma}_\alpha^{\vphantom{\dagger}}=0$.

Virtual tunneling processes can be described by the effective
Hamiltonian~\cite{kn:kaminski2000}
\begin{eqnarray}
  H_1 &=& \frac{\hbar}{4 \pi}
  {\rm P}
  \int d\xi  \sum_{\alpha} (1 - 2 f_{\alpha}(\xi))
  \nonumber \\ && \mbox{} \times
  \left[ 
  \frac{\gamma_{\alpha} \gamma_{\alpha}^{\dagger}}
  {\varepsilon_1 - \varepsilon_0 - \xi}
  +  
  \frac{\bar{\gamma}_{\alpha} \bar{\gamma}_{\alpha}^{\dagger}}
  {\varepsilon_2-\varepsilon_1 - \xi}
  \right],
  \label{eq:H1}
\end{eqnarray}
where P denotes the Cauchy principal value. Note that if 
$\varepsilon_2 - \varepsilon_1 = \varepsilon_1 -
\varepsilon_0$, one has $H_1$ proportional to the unit matrix in spin 
space and thus $\comm{H_1,\rho_1}=0$:
virtual excitations do not cause a spin
precession without interactions.~\cite{kn:koenig2003}
The time evolution of the scalars
$p_0$ and $p_2$ and the $2 \times 2$ matrix $\rho_1$ is
described by the equations\cite{kn:koenig2003,kn:braun2004}
\begin{widetext}
\begin{eqnarray}
  \label{eq:rho0}
  \ddif{p_0}{t} &=& \sum_{\alpha} 
  (1-f_{\alpha}(\varepsilon_1-\varepsilon_0))
  \gamma_{\alpha}^{\dagger} \rho_1 \gamma_{\alpha}^{\vphantom{\dagger}}
  - \sum_{\alpha} f_{\alpha}(\varepsilon_1-\varepsilon_0)
  \gamma_{\alpha}^{\dagger} p_0 \gamma_{\alpha}^{\vphantom{\dagger}}, \\
  \ddif{\rho_1}{t} &=& 
  \frac{i}{\hbar} ( \rho_1 H_1 - H_1 \rho_1) +
  \sum_{\alpha} 
  f_{\alpha}(\varepsilon_1 - \varepsilon_0)
  \gamma_{\alpha}^{\vphantom{\dagger}} p_0 \gamma_{\alpha}^{\dagger}
  - \frac{1}{2} \sum_{\alpha} 
  (1-f_\alpha(\varepsilon_1 - \varepsilon_0)) 
  \left[ \gamma_{\alpha}^{\vphantom{\dagger}} \gamma_{\alpha}^{\dagger} \rho_1
  + \rho_1 \gamma_{\alpha}^{\vphantom{\dagger}} \gamma_{\alpha}^{\dagger} \right]
  \nonumber \\ && \mbox{}
  +
  \sum_{\alpha} 
   (1 - f_{\alpha}(\varepsilon_2 - \varepsilon_1))
  \bar{\gamma}_{\alpha}^{\vphantom{\dagger}} p_2 \bar{\gamma}_{\alpha}^{\dagger}
  - \frac{1}{2} \sum_{\alpha} 
  f_\alpha(\varepsilon_2 - \varepsilon_1)
  \left[ \bar{\gamma}_{\alpha}^{\vphantom{\dagger}} \bar{\gamma}_{\alpha}^{\dagger} \rho_1
  + \rho_1 \bar{\gamma}_{\alpha}^{\vphantom{\dagger}} \bar{\gamma}_{\alpha}^{\dagger} 
  \right],  \label{eq:rho1}
\end{eqnarray}
\begin{eqnarray}
  \ddif{p_2}{t} &=&
  \sum_{\alpha} 
  f_{\alpha}(\varepsilon_2-\varepsilon_1)
  \bar{\gamma}_{\alpha}^{\dagger} \rho_1 \bar{\gamma}_{\alpha}^{\vphantom{\dagger}}
  - \sum_{\alpha} 
  (1-f_{\alpha}(\varepsilon_2-\varepsilon_1))
  \bar{\gamma}_{\alpha}^{\dagger} p_2\bar{\gamma}_{\alpha}^{\vphantom{\dagger}},
  \label{eq:rho2}
\end{eqnarray}
whereas the current through each of the tunnel contacts is calculated 
as~\cite{kn:beenakker1991,kn:averin1991}
\begin{eqnarray}
  I_{\alpha} &=&  f_{\alpha}(\varepsilon_1 - 
  \varepsilon_0) \gamma_{\alpha}^{\dagger}p_0
  \gamma_{\alpha}^{\vphantom{\dagger}} - (1 - f_{\alpha}(\varepsilon_1 - 
  \varepsilon_0)) \gamma_{\alpha}^{\dagger} \rho_1 
  \gamma_{\alpha}^{\vphantom{\dagger}}
  + f_{\alpha}(\varepsilon_2 - \varepsilon_1)
  \bar{\gamma}_{\alpha}^{\dagger} \rho_1 \bar{\gamma}_{\alpha}^{\vphantom{\dagger}}
  - (1-f_{\alpha}(\varepsilon_{2} - 
  \varepsilon_1)) \bar{\gamma}_{\alpha}^{\dagger}p_2 \bar{\gamma}_{\alpha}^{\vphantom{\dagger}}.
  \label{eq:current}
\end{eqnarray}
\end{widetext}

For $f_{\alpha}=0$ or $f_{\alpha}=1$, Eqs.~(\ref{eq:rho0})--(\ref{eq:current}) follow from
 considering the escape of
electrons or holes from the metal particle into 
vacuum.~\cite{kn:nazarov1993,kn:gurvitz1998,kn:dong2004} The factors
$f_{\alpha}$  and $(1-f_{\alpha})$, which also appear in the
scalar rate equations,~\cite{kn:beenakker1991,kn:averin1991} 
are inserted to reflect the modification of tunneling 
rates by the electron distribution in the leads. This simple
way of accounting
 for the presence of electrons in the leads
is no longer valid when correlations between the 
electrons in the leads and in the metal particle are formed,
such as  is the case in the Kondo 
effect.~\cite{kn:ng1988,kn:glazman1988,kn:ma2003,kn:martinek2003,kn:martinek2003a,kn:choi2004,kn:martinek2004} We remark that, 
since Eqs.~(\ref{eq:rho0})--(\ref{eq:current}) are meant to describe
transport at the lowest order in $\Gamma_{\alpha}$ only, the
energy shift implied by the Hamiltonian $H_1$
need not be included in the
argument of the distribution function $f_{\alpha}$. For that
reason, Eqs.~(\ref{eq:rho0})--(\ref{eq:current})
describe both the case of low bias and high bias, in contrast
to the formalism of Refs.~\onlinecite{kn:gurvitz1998,kn:dong2004}, which is appropriate
for high bias only. Also note
that Eqs.~(\ref{eq:rho0})--(\ref{eq:rho2}) are consistent
with probability conservation, Eq.~(\ref{eq:prob}), and that they
reduce to the standard rate
equations~\cite{kn:beenakker1991,kn:averin1991}
once the polarizations in the leads are collinear.

\subsection{General formalism}
\label{sec:general}
For the general description, we consider a normal-metal particle with
many-electron levels $\varepsilon_{k}$, each of which has
 degeneracy $j_k$. The
many-electron states are labeled $|k,m\rangle$, where
$m=1,\ldots,j_k$. The number of electrons in 
state $|k,m\rangle$ is denoted by $N_k$. 
Real and virtual transitions between
many-electron states are possible because of tunneling of electrons
between the metal particle and the source and drain reservoirs. As 
before, we 
assume that this tunneling rate is small in comparison to the spacing
between electronic levels and temperature. In that case, we may
describe the state of the dot by a set of density matrices $\rho_k$
for each many-electron level, and we can neglect correlations between
states with different energy. 

The tunneling Hamiltonian describing the coupling of the metal
particle to lead $\alpha$ is determined by the $j_k \times j_{k'}$
matrix $v^{\pm}_{\alpha;k,k'}$ containing the matrix elements between the
many-electron multiplets $|k,\cdot\rangle$ and $|k',\cdot\rangle$ with
$N_{k} = N_{k'} \pm 1$. In order to make contact to the rate
equations derived above, we define the $j_k \times j_{k'}$
matrix of transition amplitudes $\gamma^{\pm}_{\alpha;k,k'} =
(2 \pi \nu_{\alpha}/\hbar)^{1/2} v^{\pm}_{\alpha;k,k'}$, 
where $\nu_{\alpha}$ is the density of states of
lead $\alpha$. We define $\gamma^{\pm}_{\alpha;k,k'} = 0$ if $N_{k} \neq
N_{k'} \pm 1$. One may write $\gamma^{\pm}_{\alpha;k,k'} =
(\Gamma_{\alpha})^{1/2} w_{\alpha;k,k'}^{\pm}$, where 
$w_{\alpha;k,k'}^{\pm}$ is dimensionless and $\Gamma_{\alpha}$ is
the tunneling rate through contact $\alpha$ if the metal island is
replaced by an electron reservoir. For point contacts, the
magnitude of $w_{\alpha;k,k'}^{\pm}$ is set by the value of a 
wavefunction at the location of the contact.\cite{kn:jalabert1992}
If the degeneracy of the multiplets $|k,\cdot\rangle$
and $|k',\cdot\rangle$ arises from angular momentum, the matrix 
structure of 
$w_{\alpha;k,k'}^{\pm}$ is set by the
Clebsch-Gordon coefficients. With this notation,
virtual excitations lead to the effective 
Hamiltonian $H_k$ for the multiplet $|k,\cdot\rangle$,
\begin{eqnarray}
  H_k &=& \frac{\hbar}{4 \pi} {\rm P} \int d\xi \sum_{\alpha,k'}
  (1 - 2 f_{\alpha}(\xi))
  \nonumber \\ && \mbox{} \times
  \left[ \frac{\gamma_{\alpha;k,k'}^{+} \gamma_{\alpha;k',k}^{-}}{\varepsilon_{k} -
  \varepsilon_{k'} - \xi} +
  \frac{\gamma_{\alpha;k,k'}^{-} \gamma_{\alpha;k',k}^{+}}{\varepsilon_{k'}- \varepsilon_{k} - \xi} \right].\label{eq:precgen}
\end{eqnarray}
Then the appropriate generalization of the
rate equations (\ref{eq:rho0})--(\ref{eq:rho2}) and the current
formula (\ref{eq:current}) is
\begin{widetext}
\begin{eqnarray}
  \label{eq:rate}
  \frac{\partial \rho_k}{\partial t}
  &=& \frac{i}{\hbar} (\rho_k H_k - H_k \rho_k) +
  \sum_{\alpha,k'}  \left[ f_{\alpha}(\varepsilon_{k} -
  \varepsilon_{k'})
  \gamma^+_{\alpha;k,k'} \rho_{k'}^{\vphantom{+}} \gamma^-_{\alpha;k',k}
  + (1-f_{\alpha}(\varepsilon_{k'} - \varepsilon_{k}))
  \gamma^-_{\alpha;k,k'} \rho_{k'}^{\vphantom{+}} \gamma^+_{\alpha;k',k} \right]
  \nonumber \\
  && \mbox{} - \frac{1}{2}
  \sum_{\alpha,k}  \left[ f_{\alpha}(\varepsilon_{k'}
  - \varepsilon_{k}) (\gamma^-_{\alpha;k,k'} \gamma^+_{\alpha;k',k} \rho_k^{\vphantom{+}}
  + \rho_k^{\vphantom{+}} \gamma^-_{\alpha;k,k'}
  \gamma^+_{\alpha;k',k})
  \right. \nonumber \\ &&\qquad\qquad\qquad \ \ \ \ \ \left. \mbox{} +
  (1 - f_{\alpha}(\varepsilon_{k} - \varepsilon_{k'}))
  (\gamma^+_{\alpha;k,k'} \gamma^-_{\alpha;k',k} \rho_k^{\vphantom{+}}
  + \rho_k^{\vphantom{+}} \gamma^+_{\alpha;k,k'} \gamma^-_{\alpha;k',k}) \right],\\
  I_{\alpha} &=&
  e \sum_{k,k'}  \left[ f_{\alpha}(\varepsilon_{k} -
  \varepsilon_{k'})
  \gamma^+_{\alpha;k,k'} \rho_{k'}^{\vphantom{+}} \gamma^-_{\alpha;k',k}
  - (1-f_{\alpha}(\varepsilon_{k'} - \varepsilon_{k}))
  \gamma^-_{\alpha;k,k'} \rho_{k'}^{\vphantom{+}} \gamma^+_{\alpha;k',k} \right].
  \label{eq:currentgeneral}
\end{eqnarray}
\end{widetext}
Here the summation over $k'$ extends over all many-electron states
different from $k$. 
One easily verifies that Eq.~(\ref{eq:rate}) conserves the total
probability $\sum_{k} \mbox{tr}\, \rho_k=1$.

\subsection{Unpolarized leads}\label{sec:unpolarized}

The density-matrix rate equations (\ref{eq:rate}) and
(\ref{eq:currentgeneral}) do \emph{not} necessarily
reduce to the standard scalar
rate equations of Refs.~\onlinecite{kn:beenakker1991,kn:averin1991}
when all leads are unpolarized
or when the leads have collinear spin polarizations. 
The reason for this, at first,
surprising fact is that overlaps between different many-body states
are not accurately described by scalar transition probabilities if
there are degeneracies. With degeneracies, non-orthogonal coherent
superpositions of many-electron states are involved in the transport
process. 

Although level repulsion rules out degeneracies in the single-particle
states in a generic metal grain or quantum dot, for large
metal grains or quantum dots, (near) degeneracies may occur in the
many-electron spectrum. The origin of the degeneracy is that 
electron-electron interactions in metal particles or 
quantum dots with 
large dimensionless conductance $g$
are described by the `universal 
interaction Hamiltonian',\cite{kn:aleiner2002} 
\begin{equation}
  H_{\rm ee} = E_{\rm C} N^2 + J S^2, \label{eq:Hee}
\end{equation}
where $N$ is the total number of electrons on the metal 
particle, $E_{\rm C}$ is the charging energy, $S$ is the total 
spin, and $J$ is the exchange interaction strength. 
According to Eq.~(\ref{eq:Hee}), the energy of a many-electron 
state depends on the occupation of the single-electron states and 
the total spin $S$ only. This gives rise
to degeneracies in many-electron states with three or more
singly occupied single-electron levels. For example, in a metal
grain with single-electron levels labeled $1$, $2$, and $3$, the 
two states
\begin{subequations}
\label{eq:two}
\begin{align}
\ket{+}&\equiv \frac1{\sqrt3}\left(e^\frac{2\pi i}{3}\ccr{\spup1}\ccr{\spdown2}\ccr{\spdown3}\ket{0}+e^\frac{-2\pi i}{3}\ccr{\spdown1}\ccr{\spup2}\ccr{\spdown3}\ket{0}\right.\nonumber\\
&\qquad\qquad\qquad\left.+\ccr{\spdown1}\ccr{\spdown2}\ccr{\spup3}\ket{0}\right), \\
\ket{-}&\equiv \frac1{\sqrt3}\left(e^\frac{-2\pi i}{3}\ccr{\spup1}\ccr{\spdown2}\ccr{\spdown3}\ket{0}+e^\frac{2\pi i}{3}\ccr{\spdown1}\ccr{\spup2}\ccr{\spdown3}\ket{0}\right.\nonumber\\
&\qquad\qquad\qquad\left.+\ccr{\spdown1}\ccr{\spdown2}\ccr{\spup3}\ket{0}\right), 
\end{align}
\end{subequations}
both have three singly
occupied single-electron levels with total 
spin $S=1/2$ and $S_z = -1/2$. Hence, according to the `universal
interaction Hamiltonian', they are degenerate. Since both states
have the same value of $S_z$, the degeneracy is not
broken by a magnetic field. However, in principle
it may be lifted by non-universal
residual interactions that are not included in the `universal
interaction Hamiltonian',~\cite{kn:aleiner2002} but such residual 
interactions are weak if $g \gg 1$,
and they can be neglected if the level splitting 
that they cause is smaller 
than the level broadening due to escape through the tunnel contacts.
The degeneracy may also be lifted in metal particles with spin-orbit 
scattering if the spin-orbit rate $\hbar/\tau_{\rm so}$ is
comparable to the mean spacing $\Delta$ between single-electron 
levels.\cite{kn:gorokhov2003,kn:gorokhov2004}

We now illustrate how this degeneracy necessitates the use of a
matrix rate equation using the example of a metal particle with
three spin-degenerate single-electron levels. 
For the ease of argument, a magnetic field is applied along the 
negative $z$ axis. We consider transitions from the three 
two-electron states with $S=1$, $S_z = -1$,
\begin{subequations}
\label{eq:three}
\begin{align}
\ket{1}\equiv \ccr{\spdown2}\ccr{\spdown3}\ket{0}, \\
\ket{2}\equiv \ccr{\spdown1}\ccr{\spdown3}\ket{0}, \\
\ket{3}\equiv \ccr{\spdown1}\ccr{\spdown2}\ket{0},
\end{align}%
\end{subequations}
to the degenerate three-electron states (\ref{eq:two}). As 
the states (\ref{eq:three}) are
non-degenerate, they are described by means of the probability $p_j$
of finding the system in state $|j\rangle$, $j=1,2,3$. On the
other hand, the states (\ref{eq:two}) are degenerate and we need to 
describe their 
occupation by a $2 \times 2$
density matrix $\rho$,
\begin{equation}
  \rho = \left( \begin{array}{cc} \rho_{++} & \rho_{+-} \\
  \rho_{-+} & \rho_{--} \end{array} \right).
  \label{eq:rhodef}
\end{equation}
Transitions from the states (\ref{eq:three})
to the doublet (\ref{eq:two}) occur at rates $\Gamma_j$, $j=1,2,3$.
Writing down the time evolution of $\rho$ that results from those 
transitions, we find
\begin{align}
\frac{\ddd\rho}{\ddd t}&=\frac{\Gamma_1}{3} p_1\left(\begin{array}{cc} 1 & e^\frac{2\pi i}{3}\\ e^\frac{-2\pi i}{3} &1\end{array}\right)+\frac{\Gamma_2}{3} p_2\left(\begin{array}{cc} 1 & e^\frac{-2\pi i}{3}\\ e^\frac{2\pi i}{3} &1\end{array}\right)\nonumber \\
&\qquad+\frac{\Gamma_3}{3}p_3 \left(\begin{array}{cc} 1 & 1\\ 1
      &1\end{array}\right) + \ldots,
  \label{eq:drho}
\end{align}
where the remaining terms describe processes that do not depend on
the $p_j$, $j=1,2,3$.
Clearly, there is no basis that would diagonalize all three matrices
in Eq.~(\ref{eq:drho}) simultaneously for arbitrary choice of the
$p_j$. This proves that it is imperative to use the full matrix
structure for 
$\rho_{}$ in order to properly deal with correlations 
between the states (\ref{eq:two}). 

\section{Application to spin-polarized transport}
\label{sec:3}

We now apply the formalism of the previous section to transport
through metal particles with ferromagnetic contacts. We first consider
the simpler case of a metal particle in which only one energy level 
participates in transport, and then consider the case of multiple 
levels. 
 
\subsection{Single doubly occupied level}

The linear-response conductance $G$ of a metal particle coupled to two
fully polarized ferromagnetic leads, labeled L and R, is easily
calculated from Eqs.~(\ref{eq:rho0})-(\ref{eq:current}),
\begin{widetext}
\begin{eqnarray}
  G &=& 
  G_0 
  \cos^2(\theta/2)
  \left[1 - \frac{4 a^2 \Gamma_L \Gamma_R \sin^2(\theta/2)}
  {[a^2 + (1 - f(\varepsilon_1 -
  \varepsilon_0) + f(\varepsilon_2 - \varepsilon_1))^2]
  (\Gamma_L + \Gamma_R)^2} \right]^{-1},
  \label{condFullpol}
\end{eqnarray}
where $\theta$ is the angle between the polarizations of 
the ferromagnets, $G_0$ is the linear conductance for $\theta \to
0$,
\begin{equation}
  G_0 = \frac{e^2}{\hbar T}
  \frac{\Gamma_L \Gamma_R (1 - f(\varepsilon_2 - \varepsilon_1))
  f(\varepsilon_1 - \varepsilon_0) (1 - f(\varepsilon_1 -
  \varepsilon_0) + f(\varepsilon_2 - \varepsilon_1))}{
  (\Gamma_L + \Gamma_R)
  (1 + f(\varepsilon_1 -
  \varepsilon_0) - f(\varepsilon_2 - \varepsilon_1))},
  \label{eq:G0}
\end{equation}
and   
\begin{equation}
  a =  {\rm P}
  \int \frac{d \xi}{2 \pi}
  \frac{(1 - 2 f(\xi))
  (\varepsilon_2+\varepsilon_0-2 \varepsilon_1)}
  {(\varepsilon_1 - \varepsilon_0 - \xi)
   (\varepsilon_2-\varepsilon_1 - \xi )}.
\end{equation}

For partially polarized leads with polarization 
$P_\alpha\equiv (\Gamma_\alpha-\bar{\Gamma}_\alpha)/(\Gamma_\alpha+\bar{\Gamma}_\alpha)$, where $\bar{\Gamma}_\alpha$ is the tunneling rate for electrons with spin $\bar{m}_\alpha$, one finds
\begin{eqnarray}
  G &=&
  \frac{2 G_0}{D}
  \left[ \Gamma_L(1-P_L) + \Gamma_R(1-P_R)
  + \frac{ P_L^2 P_R^2 \Gamma_L \Gamma_R\sin^2 \theta}
  {\Gamma_L(1 + P_R) + \Gamma_R(1 +
  P_L) - D E}
  \right],
  \label{condPartpol}
\end{eqnarray}
where $G_0$ is given by Eq.~(\ref{eq:G0}) above, and we abbreviated
\begin{eqnarray*}
  D &=& \Gamma_R(1+P_L)(1-P_R) +
  \Gamma_L(1-P_L)(1+P_R) + 4 \Gamma_L \Gamma_R P_L P_R
  \sin^2(\theta/2)/(\Gamma_L + \Gamma_R), \\
  E &=& 
  a^2  (1 + P_L) (1 + P_R) (\Gamma_L + \Gamma_R)
  [a^2 +(1 - f(\varepsilon_1-\varepsilon_0) + 
  f(\varepsilon_2-\varepsilon_1))^2]^{-1}
  [\Gamma_L(1 + P_R) + \Gamma_R(1 +
  P_L)]^{-1}.
\end{eqnarray*}
\end{widetext}
For symmetric contacts, $\Gamma_L = \Gamma_R$ and $P_L = P_R$, 
Eqs.~(\ref{condFullpol}) and (\ref{condPartpol}) were previously
obtained in Refs.~\onlinecite{kn:koenig2003,kn:braun2004}.
Without the spin-precession term (the first term on the r.h.s. of
Eq.~(\ref{eq:rho1})), we recover
the linear conductance calculated by Usaj and
Baranger,~\cite{kn:usaj2001} after correction 
of a technical mistake in Ref.~\onlinecite{kn:usaj2001}.

As pointed out by K\"onig and Martinek, the role of the
spin-precession term is to reduce the angular dependence 
 of the conductance. Our general results (\ref{condFullpol}) and
(\ref{condPartpol}) show how this reduction depends on the symmetry of
the contacts: the reduction is strongest for symmetric contacts
($\Gamma_R = \Gamma_L$), whereas it vanishes in the generic case of
very asymmetric contacts ($\Gamma_R \ll \Gamma_L$). In the latter
case, the spin precession axis is aligned with $m_L$; precession
around $m_L$ does not change the angular dependence of the conductance.

\subsection{Case of up to three electrons on the dot}
\label{sec:spinApplGeneral}

For a metal particle in which more than one level contributes to
transport we calculated the differential conductance $G =
\partial I/\partial V$ numerically as a function of the 
bias voltage $V$.

The numerical calculation
was done for a metal particle in which five single-electron
levels, with a total of two or three electrons, contributed to
the current. The lead polarizations were chosen parallel or
anti-parallel, with polarizations $P_L$=$P_R\equiv P$.
With a maximum of three singly-occupied levels,
the largest possible spin on the dot 
was $3/2$. The positions of the
single-electron energy levels were taken from the center of a matrix
drawn from the Gaussian Orthogonal Ensemble of
random matrix theory, and the temperature $T$ was set at one percent
of the mean spacing $\Delta$ between the single-electron levels, to 
ensure that all features in the current-voltage characteristic could be
resolved in the numerical calculation. Electron-electron interactions
were 
 described using Eq.~(\ref{eq:Hee}). In the
numerical calculations, we set $E_{\rm C} = 25 \Delta$, and 
$J =-0.32\Delta$. (Values of the exchange constant $J$ are tabulated
in Ref.~\onlinecite{kn:gorokhov2004} for most normal metals.)
The tunneling rates $\Gamma_{L}$ and $\Gamma_{R}$ 
were chosen 
$\lesssim 0.1 k_BT$  and equal for all levels, as is appropriate for 
metal particles with wide tunnel barriers.
The source-drain voltage $V$ was
 applied to the right lead 
and was assumed to change the effective chemical potential in the right 
lead only.

The use of leads with collinear polarizations in the numerical
calculations eliminates most of the necessity of using density-matrix
rate equations, except for the degenerate $S=1/2$ states with three
singly-occupied levels. These states are fourfold degenerate. We
denote them
\begin{subequations}
\label{eq:double1}
\begin{align}
\hspace{12pt}& |\frac12,+\rangle \equiv \frac{1}{\sqrt3}(e^\frac{2\pi
    i}{3}\ket{\spup\spup\spdown}+e^\frac{-2\pi
    i}{3}\ket{\spup\spdown\spup}+\ket{\spdown\spup\spup}),
  \\
&  |\frac12,-\rangle \equiv \frac{1}{\sqrt3} (e^\frac{-2\pi
    i}{3}\ket{\spup\spup\spdown}+e^\frac{2\pi
    i}{3}\ket{\spup\spdown\spup}+\ket{\spdown\spup\spup}), 
\end{align}
\end{subequations}
\begin{subequations}
 \label{eq:double2}
\begin{align}
  |-\frac12,+\rangle &\equiv \frac{1}{\sqrt3}(e^\frac{2\pi
    i}{3}\ket{\spup\spdown\spdown}+e^\frac{-2\pi
    i}{3}\ket{\spdown\spup\spdown}+\ket{\spdown\spdown\spup}),\\
  |-\frac12,-\rangle &\equiv \frac{1}{\sqrt3}(e^\frac{-2\pi
    i}{3}\ket{\spup\spdown\spdown}+e^\frac{2\pi
    i}{3}\ket{\spdown\spup\spdown}+\ket{\spdown\spdown\spup}).
\end{align}
\end{subequations}
However, only the twofold degeneracy inside the pairs with $S_z = 1/2$ 
and $S_z = -1/2$ is relevant, and it is sufficient to describe the
occupation of the four $S=1/2$ states with two $2 \times 2$ density
matrices $\rho(S_z=1/2)$ and $\rho(S_z=-1/2)$ as in Eq.~(\ref{eq:rhodef}).

To illustrate the use of the matrix rate equations in this case, we
write down the full transition vectors for the transition between the
$S=1$ triplet states and the two doublets (\ref{eq:double1}) and
(\ref{eq:double2}). We denote the $S=1$
triplet state by $|S_z\rangle$, with $S_z = -1,0,1$. In relation to 
the two energy levels already occupied in
the triplet state, we consider adding an electron
in a single-electron level with higher, lower, or intermediate 
energy and denote the different vectors by superscripts $h$, $l$, and 
$m$, respectively. Choosing the orientation of the leads as the spin
quantization axis, the nonzero transition vectors for addition of an
electron with spin up from lead $\alpha$ as they appear in the rate 
equation (\ref{eq:rate}) are $\gamma^{+,h}_\alpha = (\Gamma^h_{\alpha})^{1/2} 
w^{+,h}$, $\gamma^{+,l}_\alpha = (\Gamma^l_{\alpha})^{1/2} w^{+,l}$, and $\gamma^{+,m}_\alpha= (\Gamma^m_{\alpha})^{1/2} w^{+,m}$, with
\begin{subequations}
\label{overlap}
\begin{eqnarray}
  w^{+,h}_{\ket{\frac12},\ket{0}}&=&
  \frac{-1}{\sqrt{6}}\left(\begin{array}{c}
  e^{-2\pi i/3}\\ e^{2\pi i/3} \end{array}\right),\\
  w^{+,h}_{\ket{-\frac12},\ket{-1}}&=&\frac1{\sqrt{3}}\left(\begin{array}{c}
  1\\ 1 \end{array}\right),\\
  w^{+,l}_{\ket{\frac12},\ket{0}}&=&\frac{-1}{\sqrt{6}}\left(\begin{array}{c}
  1\\1 \end{array}\right),\\
  w^{+,l}_{\ket{-\frac12},\ket{-1}}&=&\frac1{\sqrt{3}}\left(\begin{array}{c}
  e^{-2\pi i/3}\\ e^{2\pi i/3}\\ \end{array}\right),\\
  w^{+,m}_{\ket{\frac12},\ket{0}}&=&\frac1{\sqrt{6}}\left(\begin{array}{c}
  e^{2\pi i/3}\\e^{-2\pi i/3} \end{array}\right),\\
  w^{+,m}_{\ket{-\frac12},\ket{-1}}&=&\frac1{\sqrt{3}}\left(\begin{array}{c}
  e^{2\pi i/3}\\ e^{-2\pi i/3} \end{array}\right).
\end{eqnarray}
\end{subequations}
The transition vectors for adding a spin-down electron follow
straightforwardly from the above. The amplitudes for removing
electrons are obtained from the above by hermitian conjugation. 
Although these were not considered in the numerical calculations, 
we mention that the overlap matrices for non-collinear lead
polarizations can be obtained from Eqs.~(\ref{overlap}) by combining 
the transition vectors into 4$\times$3 matrix amplitudes for the full 
transition from the spin-1 triplet to the three-electron
spin-$\frac12$ quadruplet, followed by multiplication with appropriate 
representations of rotation matrices. In this particular case, the 
transition matrix amplitude would have to be multiplied with a 
four-dimensional representation from the left and a three-dimensional 
representation from the right. The relevant rotation matrices are
listed in the appendix.

\begin{figure}[pt]
\setlength{\unitlength}{1cm}
\includegraphics[width=8cm]{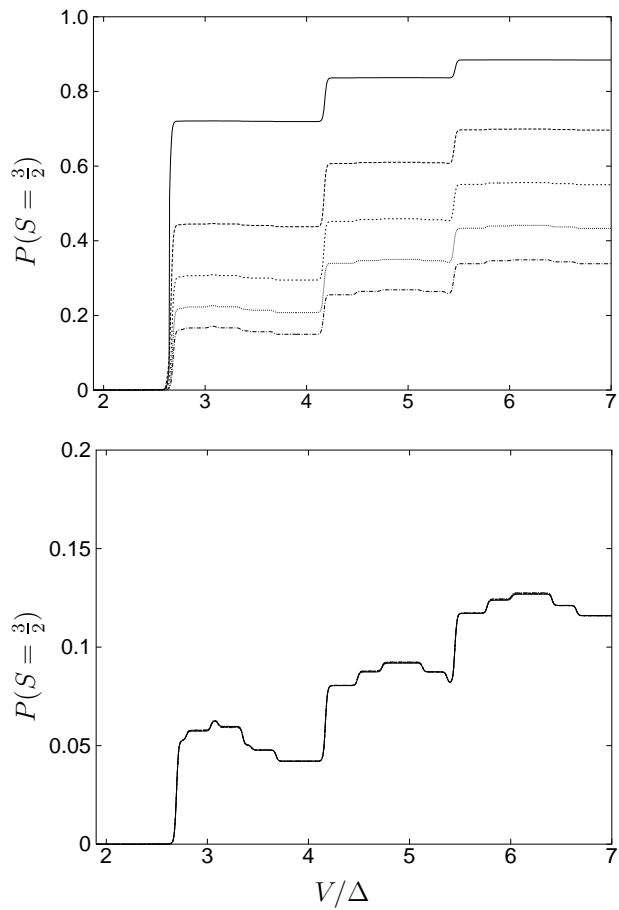}
\caption{Probability of finding total spin 3/2 on the dot for
  anti-parallel (top panel) and parallel lead polarizations 
(bottom panel) featuring polarizations $P$=$P_L$=$P_R$=0.95, 0.85, 0.75, 0.65, 0.55 (top to bottom), with $\Gamma_R/\Gamma_L=0.2$.
} \label{fig:spinAccum}
\end{figure}

\begin{figure}[pt]
\setlength{\unitlength}{1cm}
\begin{picture}
(8.5,10.75)(0.75,0)\put(0,0){\includegraphics[width=10cm]{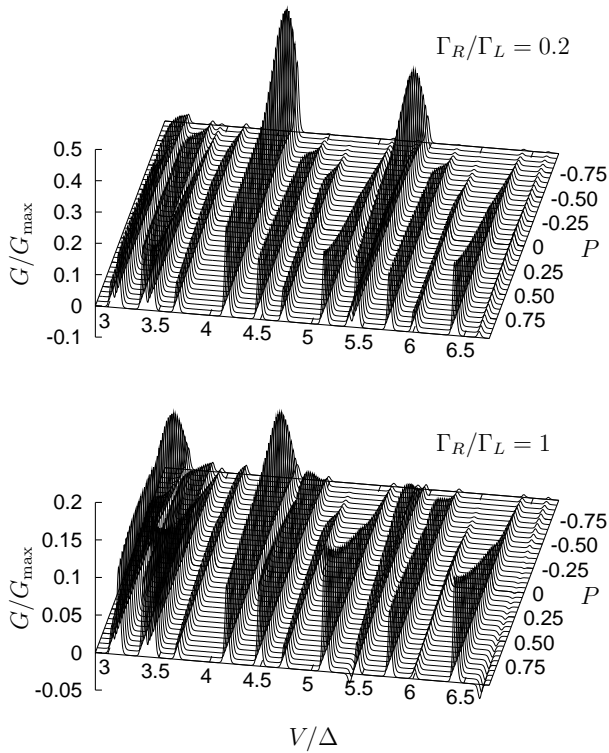}}
\end{picture}
\caption{Excerpt from the spectrum of several conductance peaks for
  anti-parallel and parallel lead orientations as a function of
  source-drain voltage $V$ (in units of the mean level spacing
  $\Delta$) and
  of lead polarization $P=P_L=P_R$. For ease of presentation, the case of parallel
  polarization is plotted against \emph{negative} polarization.
  The excerpt shown here does not
  include the dominant low-energy peak.
} \label{fig:3DPlots}
\end{figure}

\begin{figure}[pt]
\setlength{\unitlength}{1cm}
\begin{picture}
(8.5,6.5)(0,0)\put(0,0){\includegraphics[width=8cm]{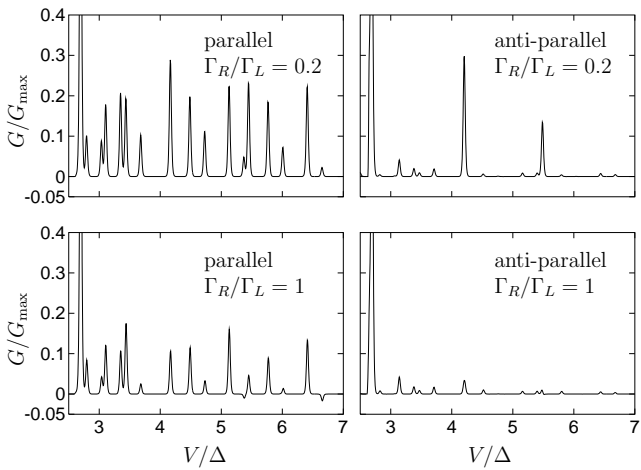}}
\end{picture}
\caption{Normalized differential conductance as a function of
  source-drain voltage $V$ for parallel (left panels) and
  anti-parallel (right panels) lead polarizations
  $P$=$P_L$=$P_R$=0.9. The vertical scale has been normalized
to the magnitude $G_{\rm max}$ of the largest (first)
conductance peak. 
} \label{fig:4Plots}
\end{figure}

We now turn to the results of our numerical calculations. One expects
that anti-parallel lead polarizations cause spin accumulation on the
metal island. This can indeed be observed in our solution of the rate
equations, as shown in Fig.~\ref{fig:spinAccum}, where we plot the
probability of finding spin $S=3/2$ (as opposed to $S=1/2$) on the
metal particle for 
different lead polarizations $P$. For anti-parallel lead polarizations,
the probability to find $S=3/2$ increases with increasing
polarization, whereas it is virtually independent of polarization for
parallel lead polarizations.

In Figs.~\ref{fig:3DPlots}~and~\ref{fig:4Plots}
we address the dependence of peaks of the
differential conductance on the lead polarization $P$. The figure
describes both parallel lead polarizations (positive $P$ in the
figure) and anti-parallel lead polarizations (negative $P$ in the
figure). Some peaks evolve non-monotonously, whereas others rise or
fall monotonously in magnitude. Of particular interest are those
conductance peaks that evolve from positive to negative
values. Negative conductance peaks at voltage $V$ arise if, upon
reaching that voltage, a many-body state that is poorly
 connected to
other states is made accessible. (This type of behavior is not limited
to ferromagnetic leads.)  Not only does there seem to be a tendency
toward negative differential conductance upon going from anti-parallel
to parallel leads but also when making the transition from
$\Gamma_R\ll\Gamma_L$ to a more symmetric coupling
$\Gamma_R\sim\Gamma_L$, as shown in Fig.~\ref{fig:4Plots}. In general,
conductance peak spectra for a different polarization orientation can
therefore look very different in terms of both position and magnitude
of the most dominant peaks, even if recorded for the same sample.

\section{Influence of spin-orbit scattering}
\label{sec:4}

The role of spin-orbit scattering inside
the metal particle is
best illustrated by considering the case of transport through one 
doubly (Kramers) degenerate level. (Spin-orbit
scattering in the reservoirs instead of in the metal island
 was considered in Ref.~\onlinecite{kn:mozyrsky2002}.)
With spin-orbit scattering, the 
two eigenfunctions of the level are spinor wavefunctions. Once the
spin quantization axes are fixed at a reference point in the metal
particle, the two spinor wavefunctions define a spatially dependent 
spin quantization axis in the metal particle. The spinors
$\spinor{m}_{\alpha}$ and $\bar{\spinor{m}}_{\alpha}$ that define
directions parallel and anti-parallel to the polarization direction
in lead $\alpha$ are defined with respect to the quantization axis
at the contact with lead $\alpha$. Hence, the presence of spin orbit 
scattering in the normal metal particle alters the spinor structure
of the transition amplitudes $\gamma_\alpha$, but it does not change the 
general structure of the rate equations. The same conclusions hold
for the general case.

The above considerations imply that, for a metal particle coupled to
ferromagnetic source and drain reservoirs via two {\em point} contacts
and with only a {\em single} level contributing to transport, the sole
effect of spin-orbit scattering is a sample-specific shift of the
angle between the polarizations in the two leads. On the other hand, for
a metal particle coupled to source and drain leads via many-channel
tunneling contacts, or for a metal particle in which more than one
level contributes to transport, the effect of spin-orbit scattering
is more complicated since different levels and different channels 
experience different rotations of the spin reference frame.
In particular, with strong spin-orbit scattering, (spin) transport
through different channels or through different levels will involve
completely different rotation angles, so that the effective degree of 
spin polarization in the junction is greatly reduced. In the limit of 
a large number of channels and strong spin-orbit scattering, the rate 
equations in fact reduce to the unpolarized case.

\section{Conclusion}
We have extended the rate-equation formalism
to the case of a normal island (metal nanoparticle, quantum dot, or single
molecule) attached to spin-polarized contacts with
non-collinear polarization directions. Our formalism provides a
transparent description of the sequential tunneling process in this
system, and is suitable for applications to both linear and nonlinear 
transport.

Whether one has to employ matrix rate equations or the simpler
scalar rate equations is determined by the symmetries and energy
degeneracies of the metal island and the leads. We distinguish 
the following cases: (i) leads and island have the same symmetries 
and degeneracies, (ii) some symmetries present on the island are 
broken in the leads, and (iii) leads and island feature the same 
symmetries but there are additional degeneracies on the dot. 
Case (i) corresponds, {\em e.g.}, to a normal-metal island with
unpolarized leads and spin-degenerate energy levels on the dot, or
to a normal-metal metal island with spin-polarized leads with collinear
polarization directions. In
that case, the standard rate equations of 
Refs.~\onlinecite{kn:beenakker1991,kn:averin1991} are applicable if 
there are no further degeneracies on the island. 
Case (i) also describes spin-polarized leads with
non-collinear polarization directions if spin degeneracy on the island
is lifted by a magnetic field, be it an applied field or the
stray field of the ferromagnetic leads. 
In contrast, the other two situations require matrix
rate equations: Case (ii) applies to normal metal particles with
ferromagnetic leads that are polarized in non-collinear directions so
that tunneling occurs into coherent superpositions of states on the
dot. The additional degeneracies required for case (iii) arise in
the many-electron spectrum of
generic metal particles with or without spin-polarized leads if the
`universal interaction Hamiltonian' describes the electron-electron
interactions on the metal particle. 
Such degeneracies can be lifted by
non-universal interaction corrections if the resulting
energy splitting is larger than the level 
broadening due to escape to the leads.
These corrections scale as $1/g$, where $g$
is the dimensionless conductance of the metal particle, and
as a consequence, the larger the size of the normal-metal island, the 
more important the use of matrix rate equations becomes.

\acknowledgments

We would like to thank  H.~U.~Baranger, K.~Flensberg, J.~Martinek, 
D.~C.~Ralph, and G.~Usaj for discussions. This work was supported by 
the Cornell Center for
Materials Research under NSF grant no.~DMR 0079992, 
by the NSF under grant no.~DMR 0334499, and by the Packard foundation.

\appendix*
\label{sec:app}
\section{Rotation matrices}
This appendix contains the rotation matrices for transformation from
the basis aligned with the quantization axis to a basis forming a
relative angle $\theta$ with the original quantization axis. We
restrict ourselves to the case of up to three electrons on the dot, so
that we only need representations up to dimensionality four. (The
four-dimensional representations correspond to the quadruplets of 
total spin $S=3/2$ and $S=1/2$ in the case of three singly-occupied
single-electron levels.) Also, we only consider polarizations in the
$xz$ plane, so that any superposition of single spins can be expressed 
in terms of real coefficients.
\subsection{Two dimensions:  spin-1/2 doublet}
In the standard basis of spin up and spin down,
\begin{align}
(1,0)\equiv \ket{\spup},\qquad (0,1)\equiv\ket{\spdown},
\end{align}
the rotation matrix is
\begin{align}
R_2=\left(\begin{array}{cc}
\cos(\theta)&\sin(\theta)\\
-\sin(\theta)&\cos(\theta)
	  \end{array}\right).
\end{align}

\subsection{Three dimensions: spin-1 triplet}
In the basis created by successively applying the spin-lowering operator $S_-$ to $\ket{\spup\spup}$,
\begin{eqnarray}
(1,0,0) &\equiv&\ket{\spup\spup}, \nonumber \\
(0,1,0) &\equiv& (\ket{\spup\spdown}+\ket{\spdown\spup})/\sqrt2, \\
(0,0,1) &\equiv& \ket{\spdown\spdown}, \nonumber
\end{eqnarray}
the rotation matrix reads
\begin{widetext}
\begin{align}
R_3=\left(\begin{array}{ccc}
\cos^2(\theta)&\sqrt2 \cos(\theta)\sin(\theta)&\sin^2(\theta)\\
-\sqrt2 \cos(\theta)\sin(\theta) &\cos^2(\theta)- \sin^2(\theta)& \sqrt2 \cos(\theta)\sin(\theta)\\
\sin^2(\theta) & -\sqrt2 \cos(\theta)\sin(\theta) & \cos^2(\theta)
	  \end{array}\right).
\end{align}

\subsection{Four dimensions: spin-3/2 quadruplet}
The maximum spin-3/2 states feature electrons in three singly occupied single-electron levels. In the basis obtained by successively applying the spin-lowering operator $S_-$ to $\ket{\spup\spup\spup}$,
\begin{align}
(1,0,0,0)&\equiv\ket{\spup\spup\spup},\qquad
(0,1,0,0)\equiv
  (\ket{\spup\spup\spdown}+\ket{\spup\spdown\spup}+\ket{\spdown\spup\spup})/\sqrt3, \nonumber \\
(0,0,1,0)&\equiv (\ket{\spup\spdown\spdown}+\ket{\spdown\spup\spdown}+\ket{\spdown\spdown\spup})/\sqrt3,\qquad
(0,0,0,1)\equiv \ket{\spdown\spdown\spdown},
\end{align}
we obtain the following rotation matrix
\begin{align}
R_{4,\frac32}= \left( \begin {array}{cccc}  
\cos^3(\theta) & \sqrt3 \cos^2(\theta) \sin(\theta) & \sqrt3 \cos(\theta) \sin^2(\theta) & \sin^3(\theta)\\
- \sqrt3 \cos^2(\theta) \sin(\theta)& \cos^3(\theta)-2\cos(\theta) \sin^2(\theta) & -\sin^3(\theta) + 2 \cos^2(\theta) \sin(\theta) & \sqrt3 \cos(\theta) \sin^2(\theta)\\
 \sqrt3 \cos(\theta) \sin^2(\theta) & \sin^3(\theta) - 2 \cos^2(\theta) \sin(\theta)& \cos^3(\theta)-2\cos(\theta) \sin^2(\theta) & \sqrt3 \cos^2(\theta) \sin(\theta) \\
-\sin^3(\theta)  & \sqrt3 \cos(\theta) \sin^2(\theta)& - \sqrt3 \cos^2(\theta) \sin(\theta) &\cos^3(\theta)
		      \end {array} \right). 
\end{align}

\subsection{Four dimensions: spin-1/2 quadruplet}
For the four-fold degenerate many-electron state with total spin
$S=1/2$ and three singly
 occupied single-electron levels, we write the 
rotation matrix in the basis
\begin{align}
(1,0,0,0)&\equiv (e^\frac{2\pi i}{3}\ket{\spup\spup\spdown}+e^\frac{-2\pi i}{3}\ket{\spup\spdown\spup}+\ket{\spdown\spup\spup})/\sqrt3,\qquad
(0,1,0,0)\equiv (e^\frac{-2\pi
    i}{3}\ket{\spup\spup\spdown}+e^\frac{2\pi
    i}{3}\ket{\spup\spdown\spup}+\ket{\spdown\spup\spup})/\sqrt3,
  \nonumber \\
(0,0,1,0)&\equiv (e^\frac{2\pi i}{3}\ket{\spup\spdown\spdown}+e^\frac{-2\pi i}{3}\ket{\spdown\spup\spdown}+\ket{\spdown\spdown\spup})/\sqrt3,\qquad
(0,0,0,1)\equiv (e^\frac{-2\pi i}{3}\ket{\spup\spdown\spdown}+e^\frac{2\pi i}{3}\ket{\spdown\spup\spdown}+\ket{\spdown\spdown\spup})/\sqrt3.
\end{align}
The first two vectors have $S_z=1/2$, the other two have $S_z = -1/2$. In
this basis, the rotation matrix then reads
\begin{align}
R_{4,\frac12}= 
\left( \begin {array}{cccc} 0& \cos \left( \theta \right) &-\frac12\sin \left( \theta \right)e^\frac{2\pi i}{3} &0\\
 \cos \left( \theta \right) &0&0&-\frac12\sin \left( \theta \right)e^\frac{-2\pi i}{3}\\
\frac12\sin \left( \theta \right)e^\frac{2\pi i}{3} &0&0&\cos(\theta)\\
0&\frac12\sin \left( \theta \right)e^\frac{-2\pi i}{3}& \cos \left( \theta \right) &0\\
\end {array} \right).
\end{align}
 
\end{widetext}


\begin{thebibliography}{49}
\expandafter\ifx\csname natexlab\endcsname\relax\def\natexlab#1{#1}\fi
\expandafter\ifx\csname bibnamefont\endcsname\relax
  \def\bibnamefont#1{#1}\fi
\expandafter\ifx\csname bibfnamefont\endcsname\relax
  \def\bibfnamefont#1{#1}\fi
\expandafter\ifx\csname citenamefont\endcsname\relax
  \def\citenamefont#1{#1}\fi
\expandafter\ifx\csname url\endcsname\relax
  \def\url#1{\texttt{#1}}\fi
\expandafter\ifx\csname urlprefix\endcsname\relax\def\urlprefix{URL }\fi
\providecommand{\bibinfo}[2]{#2}
\providecommand{\eprint}[2][]{\url{#2}}

\bibitem[{\citenamefont{\v{Z}uti\'{c} et~al.}(2004)\citenamefont{\v{Z}uti\'{c},
  Fabian, and Sarma}}]{kn:zutic2004}
\bibinfo{author}{\bibfnamefont{I.}~\bibnamefont{\v{Z}uti\'{c}}},
  \bibinfo{author}{\bibfnamefont{J.}~\bibnamefont{Fabian}}, \bibnamefont{and}
  \bibinfo{author}{\bibfnamefont{S.~D.} \bibnamefont{Sarma}},
  \bibinfo{journal}{Rev. Mod. Phys.} \textbf{\bibinfo{volume}{76}},
  \bibinfo{pages}{323} (\bibinfo{year}{2004}).

\bibitem[{\citenamefont{Berger}(1996)}]{kn:berger1996}
\bibinfo{author}{\bibfnamefont{L.}~\bibnamefont{Berger}},
  \bibinfo{journal}{Phys. Rev. B} \textbf{\bibinfo{volume}{54}},
  \bibinfo{pages}{9353} (\bibinfo{year}{1996}).

\bibitem[{\citenamefont{Slonczewski}(1996)}]{kn:slonczewski1996}
\bibinfo{author}{\bibfnamefont{J.~C.} \bibnamefont{Slonczewski}},
  \bibinfo{journal}{J. Magn. Magn. Mater.} \textbf{\bibinfo{volume}{159}},
  \bibinfo{pages}{1} (\bibinfo{year}{1996}).

\bibitem[{\citenamefont{Julli\`ere}(1975)}]{kn:julliere1975}
\bibinfo{author}{\bibfnamefont{M.}~\bibnamefont{Julli\`ere}},
  \bibinfo{journal}{Phys. Lett. A} \textbf{\bibinfo{volume}{54}},
  \bibinfo{pages}{225} (\bibinfo{year}{1975}).

\bibitem[{\citenamefont{Grabert and Devoret}(1992)}]{kn:grabert1992}
\bibinfo{editor}{\bibfnamefont{H.}~\bibnamefont{Grabert}} \bibnamefont{and}
  \bibinfo{editor}{\bibfnamefont{M.~H.} \bibnamefont{Devoret}}, eds.,
  \emph{\bibinfo{title}{Single Charge Tunneling}}, vol. \bibinfo{volume}{294}
  of \emph{\bibinfo{series}{Nato ASI Series B}} (\bibinfo{publisher}{Plenum,
  New York}, \bibinfo{year}{1992}).

\bibitem[{\citenamefont{Gu\'eron et~al.}(1999)\citenamefont{Gu\'eron, Deshmukh,
  Myers, and Ralph}}]{kn:gueron1999}
\bibinfo{author}{\bibfnamefont{S.}~\bibnamefont{Gu\'eron}},
  \bibinfo{author}{\bibfnamefont{M.~M.} \bibnamefont{Deshmukh}},
  \bibinfo{author}{\bibfnamefont{E.~B.} \bibnamefont{Myers}}, \bibnamefont{and}
  \bibinfo{author}{\bibfnamefont{D.~C.} \bibnamefont{Ralph}},
  \bibinfo{journal}{Phys. Rev. Lett.} \textbf{\bibinfo{volume}{83}},
  \bibinfo{pages}{4148} (\bibinfo{year}{1999}).

\bibitem[{\citenamefont{Deshmukh et~al.}(2001)\citenamefont{Deshmukh, Kleff,
  Gu\'eron, Bonet, Pasupathy, von Delft, and Ralph}}]{kn:deshmukh2001}
\bibinfo{author}{\bibfnamefont{M.~M.} \bibnamefont{Deshmukh}},
  \bibinfo{author}{\bibfnamefont{S.}~\bibnamefont{Kleff}},
  \bibinfo{author}{\bibfnamefont{S.}~\bibnamefont{Gu\'eron}},
  \bibinfo{author}{\bibfnamefont{E.}~\bibnamefont{Bonet}},
  \bibinfo{author}{\bibfnamefont{A.~N.} \bibnamefont{Pasupathy}},
  \bibinfo{author}{\bibfnamefont{J.}~\bibnamefont{von Delft}},
  \bibnamefont{and} \bibinfo{author}{\bibfnamefont{D.~C.} \bibnamefont{Ralph}},
  \bibinfo{journal}{Phys. Rev. Lett.} \textbf{\bibinfo{volume}{87}},
  \bibinfo{pages}{226801} (\bibinfo{year}{2001}).

\bibitem[{\citenamefont{Deshmukh and Ralph}(2002)}]{kn:deshmukh2002}
\bibinfo{author}{\bibfnamefont{M.~M.} \bibnamefont{Deshmukh}} \bibnamefont{and}
  \bibinfo{author}{\bibfnamefont{D.~C.} \bibnamefont{Ralph}},
  \bibinfo{journal}{Phys. Rev. Lett.} \textbf{\bibinfo{volume}{89}},
  \bibinfo{pages}{266803} (\bibinfo{year}{2002}).

\bibitem[{\citenamefont{Potok et~al.}(2003)\citenamefont{Potok, Folk, Marcus,
  Umansky, Hanson, , and Gossard}}]{kn:potok2003}
\bibinfo{author}{\bibfnamefont{R.~M.} \bibnamefont{Potok}},
  \bibinfo{author}{\bibfnamefont{J.~A.} \bibnamefont{Folk}},
  \bibinfo{author}{\bibfnamefont{C.~M.} \bibnamefont{Marcus}},
  \bibinfo{author}{\bibfnamefont{V.}~\bibnamefont{Umansky}},
  \bibinfo{author}{\bibfnamefont{M.}~\bibnamefont{Hanson}}, , \bibnamefont{and}
  \bibinfo{author}{\bibfnamefont{A.~C.} \bibnamefont{Gossard}},
  \bibinfo{journal}{Phys. Rev. Lett.} \textbf{\bibinfo{volume}{91}},
  \bibinfo{pages}{016802} (\bibinfo{year}{2003}).

\bibitem[{\citenamefont{Pasupathy et~al.}(2004)\citenamefont{Pasupathy,
  Bialczak, Martinek, Grose, Donev, McEuen, and Ralph}}]{kn:pasupathy2004}
\bibinfo{author}{\bibfnamefont{A.}~\bibnamefont{Pasupathy}},
  \bibinfo{author}{\bibfnamefont{R.}~\bibnamefont{Bialczak}},
  \bibinfo{author}{\bibfnamefont{J.}~\bibnamefont{Martinek}},
  \bibinfo{author}{\bibfnamefont{J.}~\bibnamefont{Grose}},
  \bibinfo{author}{\bibfnamefont{L.}~\bibnamefont{Donev}},
  \bibinfo{author}{\bibfnamefont{P.}~\bibnamefont{McEuen}}, \bibnamefont{and}
  \bibinfo{author}{\bibfnamefont{D.~C.} \bibnamefont{Ralph}},
  \bibinfo{journal}{preprint}  (\bibinfo{year}{2004}).

\bibitem[{\citenamefont{Barna\'s and Fert}(1998)}]{kn:barnas1998}
\bibinfo{author}{\bibfnamefont{J.}~\bibnamefont{Barna\'s}} \bibnamefont{and}
  \bibinfo{author}{\bibfnamefont{A.}~\bibnamefont{Fert}},
  \bibinfo{journal}{Phys. Rev. Lett.} \textbf{\bibinfo{volume}{80}},
  \bibinfo{pages}{1058} (\bibinfo{year}{1998}).

\bibitem[{\citenamefont{Takahashi and Maekawa}(1998)}]{kn:takahashi1998}
\bibinfo{author}{\bibfnamefont{S.}~\bibnamefont{Takahashi}} \bibnamefont{and}
  \bibinfo{author}{\bibfnamefont{S.}~\bibnamefont{Maekawa}},
  \bibinfo{journal}{Phys. Rev. Lett} \textbf{\bibinfo{volume}{80}},
  \bibinfo{pages}{1758} (\bibinfo{year}{1998}).

\bibitem[{\citenamefont{Brataas et~al.}(1999)\citenamefont{Brataas, Nazarov,
  Inoue, and Bauer}}]{kn:brataas1999}
\bibinfo{author}{\bibfnamefont{A.}~\bibnamefont{Brataas}},
  \bibinfo{author}{\bibfnamefont{Y.~V.} \bibnamefont{Nazarov}},
  \bibinfo{author}{\bibfnamefont{J.}~\bibnamefont{Inoue}}, \bibnamefont{and}
  \bibinfo{author}{\bibfnamefont{G.~E.~W.} \bibnamefont{Bauer}},
  \bibinfo{journal}{Phys. Rev. B} \textbf{\bibinfo{volume}{59}},
  \bibinfo{pages}{93} (\bibinfo{year}{1999}).

\bibitem[{\citenamefont{Barna\'s et~al.}(2000)\citenamefont{Barna\'s, Martinek,
  Michalek, Bulka, and Fert}}]{kn:barnas2000}
\bibinfo{author}{\bibfnamefont{J.}~\bibnamefont{Barna\'s}},
  \bibinfo{author}{\bibfnamefont{J.}~\bibnamefont{Martinek}},
  \bibinfo{author}{\bibfnamefont{G.}~\bibnamefont{Michalek}},
  \bibinfo{author}{\bibfnamefont{B.~R.} \bibnamefont{Bulka}}, \bibnamefont{and}
  \bibinfo{author}{\bibfnamefont{A.}~\bibnamefont{Fert}},
  \bibinfo{journal}{Phys. Rev. B} \textbf{\bibinfo{volume}{62}},
  \bibinfo{pages}{12363} (\bibinfo{year}{2000}).

\bibitem[{\citenamefont{Rudzinski and Barna\'s}(2001)}]{kn:rudzinski2001}
\bibinfo{author}{\bibfnamefont{W.}~\bibnamefont{Rudzinski}} \bibnamefont{and}
  \bibinfo{author}{\bibfnamefont{J.}~\bibnamefont{Barna\'s}},
  \bibinfo{journal}{Phys. Rev. B} \textbf{\bibinfo{volume}{64}},
  \bibinfo{pages}{085318} (\bibinfo{year}{2001}).

\bibitem[{\citenamefont{Usaj and Baranger}(2001)}]{kn:usaj2001}
\bibinfo{author}{\bibfnamefont{G.}~\bibnamefont{Usaj}} \bibnamefont{and}
  \bibinfo{author}{\bibfnamefont{H.~U.} \bibnamefont{Baranger}},
  \bibinfo{journal}{Phys. Rev. B} \textbf{\bibinfo{volume}{63}},
  \bibinfo{pages}{184418} (\bibinfo{year}{2001}).

\bibitem[{\citenamefont{Martinek et~al.}(2002)\citenamefont{Martinek, Barna\'s,
  Maekawa, Schoeller, and Sch\"on}}]{kn:martinek2002}
\bibinfo{author}{\bibfnamefont{J.}~\bibnamefont{Martinek}},
  \bibinfo{author}{\bibfnamefont{J.}~\bibnamefont{Barna\'s}},
  \bibinfo{author}{\bibfnamefont{S.}~\bibnamefont{Maekawa}},
  \bibinfo{author}{\bibfnamefont{H.}~\bibnamefont{Schoeller}},
  \bibnamefont{and} \bibinfo{author}{\bibfnamefont{G.}~\bibnamefont{Sch\"on}},
  \bibinfo{journal}{Phys. Rev. B} \textbf{\bibinfo{volume}{66}},
  \bibinfo{pages}{014402} (\bibinfo{year}{2002}).

\bibitem[{\citenamefont{Sergueev et~al.}(2002)\citenamefont{Sergueev, feng Sun,
  Guo, Wang, and Wang}}]{kn:sergueev2002}
\bibinfo{author}{\bibfnamefont{N.}~\bibnamefont{Sergueev}},
  \bibinfo{author}{\bibfnamefont{Q.}~\bibnamefont{feng Sun}},
  \bibinfo{author}{\bibfnamefont{H.}~\bibnamefont{Guo}},
  \bibinfo{author}{\bibfnamefont{B.~G.} \bibnamefont{Wang}}, \bibnamefont{and}
  \bibinfo{author}{\bibfnamefont{J.}~\bibnamefont{Wang}},
  \bibinfo{journal}{Phys. Rev. B} \textbf{\bibinfo{volume}{65}},
  \bibinfo{pages}{165303} (\bibinfo{year}{2002}).

\bibitem[{\citenamefont{Fransson et~al.}(2002)\citenamefont{Fransson, Eriksson,
  and Sandalov}}]{kn:fransson2002}
\bibinfo{author}{\bibfnamefont{J.}~\bibnamefont{Fransson}},
  \bibinfo{author}{\bibfnamefont{O.}~\bibnamefont{Eriksson}}, \bibnamefont{and}
  \bibinfo{author}{\bibfnamefont{I.}~\bibnamefont{Sandalov}},
  \bibinfo{journal}{Phys. Rev. Lett.} \textbf{\bibinfo{volume}{88}},
  \bibinfo{pages}{226601} (\bibinfo{year}{2002}).

\bibitem[{\citenamefont{Cottet et~al.}(2004{\natexlab{a}})\citenamefont{Cottet,
  Belzig, and Bruder}}]{kn:cottet2004}
\bibinfo{author}{\bibfnamefont{A.}~\bibnamefont{Cottet}},
  \bibinfo{author}{\bibfnamefont{W.}~\bibnamefont{Belzig}}, \bibnamefont{and}
  \bibinfo{author}{\bibfnamefont{C.}~\bibnamefont{Bruder}},
  \bibinfo{journal}{Phys. Rev. Lett.} \textbf{\bibinfo{volume}{92}},
  \bibinfo{pages}{206801} (\bibinfo{year}{2004}{\natexlab{a}}); \bibinfo{author}{\bibfnamefont{A.}~\bibnamefont{Cottet}},
  \bibinfo{author}{\bibfnamefont{W.}~\bibnamefont{Belzig}}, \bibnamefont{and}
  \bibinfo{author}{\bibfnamefont{C.}~\bibnamefont{Bruder}},
  \bibinfo{journal}{Phys. Rev. B} \textbf{\bibinfo{volume}{70}},
  \bibinfo{pages}{115315} (\bibinfo{year}{2004}{\natexlab{b}}).


\bibitem[{\citenamefont{Cottet and Belzig}(2004)}]{kn:cottet2004b}
\bibinfo{author}{\bibfnamefont{A.}~\bibnamefont{Cottet}} \bibnamefont{and}
  \bibinfo{author}{\bibfnamefont{W.}~\bibnamefont{Belzig}},
  \bibinfo{journal}{Europhys. Lett.} \textbf{\bibinfo{volume}{66}},
  \bibinfo{pages}{405} (\bibinfo{year}{2004}).

\bibitem[{\citenamefont{Ma and Lei}(2004)}]{kn:ma2003}
\bibinfo{author}{\bibfnamefont{J.}~\bibnamefont{Ma}} \bibnamefont{and}
  \bibinfo{author}{\bibfnamefont{X.~L.} \bibnamefont{Lei}},
  \bibinfo{journal}{Europhys. Lett} \textbf{\bibinfo{volume}{67}},
  \bibinfo{pages}{432} (\bibinfo{year}{2004}).

\bibitem[{\citenamefont{Martinek
  et~al.}(2003{\natexlab{a}})\citenamefont{Martinek, Utsumi, Imamura, Barna\'s,
  Maekawa, K\"onig, and Sch\"on}}]{kn:martinek2003}
\bibinfo{author}{\bibfnamefont{J.}~\bibnamefont{Martinek}},
  \bibinfo{author}{\bibfnamefont{Y.}~\bibnamefont{Utsumi}},
  \bibinfo{author}{\bibfnamefont{H.}~\bibnamefont{Imamura}},
  \bibinfo{author}{\bibfnamefont{J.}~\bibnamefont{Barna\'s}},
  \bibinfo{author}{\bibfnamefont{S.}~\bibnamefont{Maekawa}},
  \bibinfo{author}{\bibfnamefont{J.}~\bibnamefont{K\"onig}}, \bibnamefont{and}
  \bibinfo{author}{\bibfnamefont{G.}~\bibnamefont{Sch\"on}},
  \bibinfo{journal}{Phys. Rev. Lett.} \textbf{\bibinfo{volume}{91}},
  \bibinfo{pages}{127203} (\bibinfo{year}{2003}{\natexlab{a}}).

\bibitem[{\citenamefont{Martinek
  et~al.}(2003{\natexlab{b}})\citenamefont{Martinek, Sindel, Borda, Barna\'s,
  K\"onig, Sch\"on, and von Delft}}]{kn:martinek2003a}
\bibinfo{author}{\bibfnamefont{J.}~\bibnamefont{Martinek}},
  \bibinfo{author}{\bibfnamefont{M.}~\bibnamefont{Sindel}},
  \bibinfo{author}{\bibfnamefont{L.}~\bibnamefont{Borda}},
  \bibinfo{author}{\bibfnamefont{J.}~\bibnamefont{Barna\'s}},
  \bibinfo{author}{\bibfnamefont{J.}~\bibnamefont{K\"onig}},
  \bibinfo{author}{\bibfnamefont{G.}~\bibnamefont{Sch\"on}}, \bibnamefont{and}
  \bibinfo{author}{\bibfnamefont{J.}~\bibnamefont{von Delft}},
  \bibinfo{journal}{Phys. Rev. Lett.} \textbf{\bibinfo{volume}{91}},
  \bibinfo{pages}{247202} (\bibinfo{year}{2003}{\natexlab{b}}).

\bibitem[{\citenamefont{Choi et~al.}(2004)\citenamefont{Choi, S\'anchez, and
  L\'opez}}]{kn:choi2004}
\bibinfo{author}{\bibfnamefont{M.-S.} \bibnamefont{Choi}},
  \bibinfo{author}{\bibfnamefont{D.}~\bibnamefont{S\'anchez}},
  \bibnamefont{and} \bibinfo{author}{\bibfnamefont{R.}~\bibnamefont{L\'opez}},
  \bibinfo{journal}{Phys. Rev. Lett. 92, 056601 (2004)}
  \textbf{\bibinfo{volume}{92}}, \bibinfo{pages}{056601}
  (\bibinfo{year}{2004}).

\bibitem[{\citenamefont{Martinek et~al.}(2004)\citenamefont{Martinek, Sindel,
  Borda, Barna\'s, Bulla, K\"onig, Sch\"on, Maekawa, and von
  Delft}}]{kn:martinek2004}
\bibinfo{author}{\bibfnamefont{J.}~\bibnamefont{Martinek}},
  \bibinfo{author}{\bibfnamefont{M.}~\bibnamefont{Sindel}},
  \bibinfo{author}{\bibfnamefont{L.}~\bibnamefont{Borda}},
  \bibinfo{author}{\bibfnamefont{J.}~\bibnamefont{Barna\'s}},
  \bibinfo{author}{\bibfnamefont{R.}~\bibnamefont{Bulla}},
  \bibinfo{author}{\bibfnamefont{J.}~\bibnamefont{K\"onig}},
  \bibinfo{author}{\bibfnamefont{G.}~\bibnamefont{Sch\"on}},
  \bibinfo{author}{\bibfnamefont{S.}~\bibnamefont{Maekawa}}, \bibnamefont{and}
  \bibinfo{author}{\bibfnamefont{J.}~\bibnamefont{von Delft}},
  \bibinfo{journal}{cond-mat/0406323}  (\bibinfo{year}{2004}).

\bibitem[{\citenamefont{Pedersen et~al.}(2004)\citenamefont{Pedersen,
  Thomassen, and Flensberg}}]{kn:pedersen2004}
\bibinfo{author}{\bibfnamefont{J.~N.} \bibnamefont{Pedersen}},
  \bibinfo{author}{\bibfnamefont{J.~Q.} \bibnamefont{Thomassen}},
  \bibnamefont{and}
  \bibinfo{author}{\bibfnamefont{K.}~\bibnamefont{Flensberg}},
  \bibinfo{journal}{cond-mat}  (\bibinfo{year}{2004}).

\bibitem[{\citenamefont{Waintal and Brouwer}(2003)}]{kn:waintal2003}
\bibinfo{author}{\bibfnamefont{X.}~\bibnamefont{Waintal}} \bibnamefont{and}
  \bibinfo{author}{\bibfnamefont{P.~W.} \bibnamefont{Brouwer}},
  \bibinfo{journal}{Phys. Rev. Lett.} \textbf{\bibinfo{volume}{91}},
  \bibinfo{pages}{247201} (\bibinfo{year}{2003}).

\bibitem[{\citenamefont{K\"onig and Martinek}(2003)}]{kn:koenig2003}
\bibinfo{author}{\bibfnamefont{J.}~\bibnamefont{K\"onig}} \bibnamefont{and}
  \bibinfo{author}{\bibfnamefont{J.}~\bibnamefont{Martinek}},
  \bibinfo{journal}{Phys. Rev. Lett.} \textbf{\bibinfo{volume}{90}},
  \bibinfo{pages}{166602} (\bibinfo{year}{2003}).

\bibitem[{\citenamefont{Braun et~al.}(2004)\citenamefont{Braun, K\"onig, and
  Martinek}}]{kn:braun2004}
\bibinfo{author}{\bibfnamefont{M.}~\bibnamefont{Braun}},
  \bibinfo{author}{\bibfnamefont{J.}~\bibnamefont{K\"onig}}, \bibnamefont{and}
  \bibinfo{author}{\bibfnamefont{J.}~\bibnamefont{Martinek}},
  \bibinfo{journal}{Phys. Rev. B} \textbf{\bibinfo{volume}{70}},
  \bibinfo{pages}{195345} (\bibinfo{year}{2004}).

\bibitem[{\citenamefont{Shekhter}(1972)}]{kn:shekhter1972}
\bibinfo{author}{\bibfnamefont{R.~I.} \bibnamefont{Shekhter}},
  \bibinfo{journal}{Zh. Eksp. Teor. Fiz.} \textbf{\bibinfo{volume}{63}},
  \bibinfo{pages}{1410} (\bibinfo{year}{1972}), \bibinfo{note}{[Sov. Phys. JETP
  {\bf 36}, 747 (1973)]}.

\bibitem[{\citenamefont{Kulik and Shekhter}(1975)}]{kn:kulik1975}
\bibinfo{author}{\bibfnamefont{I.~O.} \bibnamefont{Kulik}} \bibnamefont{and}
  \bibinfo{author}{\bibfnamefont{R.~I.} \bibnamefont{Shekhter}},
  \bibinfo{journal}{Zh. Eksp. Teor. Fiz.} \textbf{\bibinfo{volume}{68}},
  \bibinfo{pages}{623} (\bibinfo{year}{1975}), \bibinfo{note}{[Sov. Phys. JETP
  {\bf 41}, 308 (1975)]}.

\bibitem[{\citenamefont{Averin and Likharev}(1986)}]{kn:averin1986}
\bibinfo{author}{\bibfnamefont{D.~V.} \bibnamefont{Averin}} \bibnamefont{and}
  \bibinfo{author}{\bibfnamefont{K.~K.} \bibnamefont{Likharev}},
  \bibinfo{journal}{J. Low Temp. Phys.} \textbf{\bibinfo{volume}{62}},
  \bibinfo{pages}{345} (\bibinfo{year}{1986}).

\bibitem[{\citenamefont{Beenakker}(1991)}]{kn:beenakker1991}
\bibinfo{author}{\bibfnamefont{C.~W.~J.} \bibnamefont{Beenakker}},
  \bibinfo{journal}{Phys. Rev. B} \textbf{\bibinfo{volume}{44}},
  \bibinfo{pages}{1646} (\bibinfo{year}{1991}).

\bibitem[{\citenamefont{Averin et~al.}(1991)\citenamefont{Averin, Korotkhov,
  and Likharev}}]{kn:averin1991}
\bibinfo{author}{\bibfnamefont{D.~V.} \bibnamefont{Averin}},
  \bibinfo{author}{\bibfnamefont{A.~N.} \bibnamefont{Korotkhov}},
  \bibnamefont{and} \bibinfo{author}{\bibfnamefont{K.~K.}
  \bibnamefont{Likharev}}, \bibinfo{journal}{Phys.\ Rev.\ B}
  \textbf{\bibinfo{volume}{44}}, \bibinfo{pages}{6199} (\bibinfo{year}{1991}).

\bibitem[{\citenamefont{\mbox{Yu.} V.~Nazarov}(1993)}]{kn:nazarov1993}
\bibinfo{author}{\bibnamefont{\mbox{Yu.} V.~Nazarov}},
  \bibinfo{journal}{Physica B} \textbf{\bibinfo{volume}{189}},
  \bibinfo{pages}{57} (\bibinfo{year}{1993}).

\bibitem[{\citenamefont{Gurvitz}(1998)}]{kn:gurvitz1998}
\bibinfo{author}{\bibfnamefont{S.~A.} \bibnamefont{Gurvitz}},
  \bibinfo{journal}{Phys. Rev. B} \textbf{\bibinfo{volume}{57}},
  \bibinfo{pages}{6602} (\bibinfo{year}{1998}).

\bibitem[{\citenamefont{von Delft and Ralph}(2001)}]{kn:vondelft2001}
\bibinfo{author}{\bibfnamefont{J.}~\bibnamefont{von Delft}} \bibnamefont{and}
  \bibinfo{author}{\bibfnamefont{D.~C.} \bibnamefont{Ralph}},
  \bibinfo{journal}{Phys. Rep.} \textbf{\bibinfo{volume}{345}},
  \bibinfo{pages}{61} (\bibinfo{year}{2001}).

\bibitem[{\citenamefont{Aleiner et~al.}(2002)\citenamefont{Aleiner, Brouwer,
  and Glazman}}]{kn:aleiner2002}
\bibinfo{author}{\bibfnamefont{I.~L.} \bibnamefont{Aleiner}},
  \bibinfo{author}{\bibfnamefont{P.~W.} \bibnamefont{Brouwer}},
  \bibnamefont{and} \bibinfo{author}{\bibfnamefont{L.~I.}
  \bibnamefont{Glazman}}, \bibinfo{journal}{Phys. Rep.}
  \textbf{\bibinfo{volume}{358}}, \bibinfo{pages}{309} (\bibinfo{year}{2002}).

\bibitem[{\citenamefont{Alhassid et~al.}(2002)\citenamefont{Alhassid, Rupp,
  Kaminski, and Glazman}}]{kn:alhassid2002}
\bibinfo{author}{\bibfnamefont{Y.}~\bibnamefont{Alhassid}},
  \bibinfo{author}{\bibfnamefont{T.}~\bibnamefont{Rupp}},
  \bibinfo{author}{\bibfnamefont{A.}~\bibnamefont{Kaminski}}, \bibnamefont{and}
  \bibinfo{author}{\bibfnamefont{L.~I.} \bibnamefont{Glazman}},
  \bibinfo{journal}{Phys. Rev. B} \textbf{\bibinfo{volume}{69}},
  \bibinfo{pages}{115331} (\bibinfo{year}{2002}).

\bibitem[{\citenamefont{Kaminski and Glazman}(2000)}]{kn:kaminski2000}
\bibinfo{author}{\bibfnamefont{A.}~\bibnamefont{Kaminski}} \bibnamefont{and}
  \bibinfo{author}{\bibfnamefont{L.~I.} \bibnamefont{Glazman}},
  \bibinfo{journal}{Phys. Rev. B} \textbf{\bibinfo{volume}{61}},
  \bibinfo{pages}{15927} (\bibinfo{year}{2000}).

\bibitem[{\citenamefont{Dong et~al.}(2004)\citenamefont{Dong, Cui, and
  Lei}}]{kn:dong2004}
\bibinfo{author}{\bibfnamefont{B.}~\bibnamefont{Dong}},
  \bibinfo{author}{\bibfnamefont{H.~L.} \bibnamefont{Cui}}, \bibnamefont{and}
  \bibinfo{author}{\bibfnamefont{X.~L.} \bibnamefont{Lei}},
  \bibinfo{journal}{Phys. Rev. B} \textbf{\bibinfo{volume}{69}},
  \bibinfo{pages}{035324} (\bibinfo{year}{2004}).

\bibitem[{\citenamefont{Ng and Lee}(1988)}]{kn:ng1988}
\bibinfo{author}{\bibfnamefont{T.~K.} \bibnamefont{Ng}} \bibnamefont{and}
  \bibinfo{author}{\bibfnamefont{P.~A.} \bibnamefont{Lee}},
  \bibinfo{journal}{Phys. Rev. Lett.} \textbf{\bibinfo{volume}{61}},
  \bibinfo{pages}{1768} (\bibinfo{year}{1988}).

\bibitem[{\citenamefont{Glazman and Raikh}(1988)}]{kn:glazman1988}
\bibinfo{author}{\bibfnamefont{L.~I.} \bibnamefont{Glazman}} \bibnamefont{and}
  \bibinfo{author}{\bibfnamefont{M.~E.} \bibnamefont{Raikh}},
  \bibinfo{journal}{Pis'ma Zh. Eksp. Teor. Fiz.} \textbf{\bibinfo{volume}{47}},
  \bibinfo{pages}{378} (\bibinfo{year}{1988}), \bibinfo{note}{[JETP Lett.\ {\bf
  47}, 452 (1988)]}.

\bibitem[{\citenamefont{Jalabert et~al.}(1992)\citenamefont{Jalabert, Stone,
  and Alhassid}}]{kn:jalabert1992}
\bibinfo{author}{\bibfnamefont{R.~A.} \bibnamefont{Jalabert}},
  \bibinfo{author}{\bibfnamefont{A.~D.} \bibnamefont{Stone}}, \bibnamefont{and}
  \bibinfo{author}{\bibfnamefont{Y.}~\bibnamefont{Alhassid}},
  \bibinfo{journal}{Phys. Rev. Lett.} \textbf{\bibinfo{volume}{68}},
  \bibinfo{pages}{3468} (\bibinfo{year}{1992}).

\bibitem[{\citenamefont{Gorokhov and Brouwer}(2003)}]{kn:gorokhov2003}
\bibinfo{author}{\bibfnamefont{D.~A.} \bibnamefont{Gorokhov}} \bibnamefont{and}
  \bibinfo{author}{\bibfnamefont{P.~W.} \bibnamefont{Brouwer}},
  \bibinfo{journal}{Phys. Rev. Lett} \textbf{\bibinfo{volume}{91}},
  \bibinfo{pages}{186602} (\bibinfo{year}{2003}).

\bibitem[{\citenamefont{Gorokhov and Brouwer}(2004)}]{kn:gorokhov2004}
\bibinfo{author}{\bibfnamefont{D.~A.} \bibnamefont{Gorokhov}} \bibnamefont{and}
  \bibinfo{author}{\bibfnamefont{P.~W.} \bibnamefont{Brouwer}},
  \bibinfo{journal}{Phys. Rev. B} \textbf{\bibinfo{volume}{69}},
  \bibinfo{pages}{155417} (\bibinfo{year}{2004}).

\bibitem[{\citenamefont{Mozyrsky et~al.}(2002)\citenamefont{Mozyrsky,
  Fedichkin, Gurvitz, and Berman}}]{kn:mozyrsky2002}
\bibinfo{author}{\bibfnamefont{D.}~\bibnamefont{Mozyrsky}},
  \bibinfo{author}{\bibfnamefont{L.}~\bibnamefont{Fedichkin}},
  \bibinfo{author}{\bibfnamefont{S.~A.} \bibnamefont{Gurvitz}},
  \bibnamefont{and} \bibinfo{author}{\bibfnamefont{G.~P.}
  \bibnamefont{Berman}}, \bibinfo{journal}{Phys. Rev. B}
  \textbf{\bibinfo{volume}{66}}, \bibinfo{pages}{161313}
  (\bibinfo{year}{2002}).

\end{thebibliography}
\end{document}